\documentclass{elsart}%
\usepackage{amsfonts}
\usepackage{natbib}
\usepackage{graphicx}
\usepackage[latin1]{inputenc}

\usepackage{amsmath}
\usepackage{amssymb}
\graphicspath{{perez-fig/}}

\newtheorem{theorem}{Theorem}
\begin{document}
\begin{frontmatter}
\title{Gravity, Dimension, Equilibrium, \& Thermodynamics}
\author{Jér\^{o}me Perez}
\address{Laboratoire de Mathématiques Appliquées, Ecole Nationale Supérieure de Techniques Avancées, 15 Bd Victor, 75739 Paris Cedex 15}
\ead{jerome.perez@ensta.fr}
\begin{abstract}
Is it actually possible to interpret gravitation as space's property in a pure classical way.  Then, we note that extended self-gravitating
system equilibrium depends directly on the number of dimension of the space in which it evolves. Given those precisions,
we review the principal thermodynamical knowledge in the context of classical gravity with arbitrary dimension of space.
Stability analyses for bounded 3D systems, namely the Antonov instability paradigm, are then rapproched to some amazing properties of
globular clusters and galaxies.
\end{abstract}
\end{frontmatter}

\section{Nature of classical gravitation}
\subsection{Gravitational field}
In a classical way, gravitation is a force $F=-\mathrm{grad}U$ acting on a
test mass $m$, deriving from a scalar potential energy $U$ which at any time
$t$ depends only on the position $\mathbf{r}\in\mathbb{R}^{3}$. This potential
is generated by a mass density field $\rho\left(  \mathbf{r},t\right)  $ such
that Poisson equation holds
\begin{equation}
U(\mathbf{r},t)=m\psi(\mathbf{r},t)\;\;\text{and}\;\;\Delta_{3}\psi\left(
\mathbf{r},t\right)  =4\pi G\rho\left(  \mathbf{r},t\right)
\end{equation}
Inverting the laplacian $\Delta_{3}$, one writes the gravitational potential
as
\begin{equation}
\psi\left(  \mathbf{r},t\right)  =-G\int\frac{\rho\left(  \mathbf{r}^{\prime
},t\right)  }{\left\vert \mathbf{r-r}^{\prime}\right\vert }d\mathbf{r}%
^{\prime}\label{conv}%
\end{equation}
Using a convolution product, this last relation writes
\begin{equation}
\psi=4\pi\ G\ g_{3}\ast\rho\label{poisconv}%
\end{equation}
Where $g_{3}$ is a solution of the equation
\begin{equation}
\Delta_{3}\ g_{3}=\delta\text{ \ ,}%
\end{equation}
which is different from harmonic polynomials. In other words $g_{3}$ is the
laplacian Green's function in $\mathbb{R}^{3}$
\begin{equation}
g_{3}\left(  \mathbf{r}\right)  =-\frac{1}{4\pi}\frac{1}{\left\vert
\mathbf{r}\right\vert }%
\end{equation}

Using information theory, interpretation of equation $\left(
\text{\ref{poisconv}}\right)  $ is clear : Green's function is
interpretable as impulsional response of the corresponding operator,
gravitational potential $\psi\left(  \mathbf{r},t\right)  $ is then the
response from space when it is submitted to the presence of a mass density
distribution $\rho\left(  \mathbf{r},t\right)  $. Constant factors are also
clearly interpretable : the Newton-Cavendish constant $G$ fixes units and
$4\pi$ is the value of the surface of unit radius sphere in $\mathbb{R}^{3}$,
it is then attached to the radial nature of the gravitationnal interaction.

Let us note that the traditional link usually made between gravity and space
in general relativity is already contained, using this formulation, in
classical field gravity. Two main problems remain: the instantaneous
character of the interaction and the fundamental restriction to isotropic
spaces (laplacians acts equally in all directions). Two major fulfillments of
Einstein theory of gravitation were to solve these problems.

In a classical context, one can directly generalize equation $\left(
\text{\ref{poisconv}}\right)  $ to obtain the classical definition of the
gravitational potential in a $n-$dimensionnal space, we have
\begin{equation}
\psi=G\ S_{n-1}\ g_{n}\ast\rho\label{ngrav}%
\end{equation}

where
\[
S_{n-1}=\frac{2\pi^{n/2}}{\Gamma(n/2)}\ \ \ \text{with\ \ \ \ }\forall
z\in\mathbb{R}_{\ast}^{+}\;\;\;\Gamma\left(  z\right)  :=\int_{0}^{+\infty
}e^{-s}s^{z-1}ds
\]
represents the value of the surface of unit radius sphere in $\mathbb{R}^{n}$,
and
\[
\forall\mathbf{r}\in\mathbb{R}_{\ast}^{n}\ \ \ \ \ \ g_{n}\left(
\mathbf{r}\right)  =\left[
\begin{array}
[c]{lc}%
\left\vert \mathbf{r}\right\vert  & \text{if }n=1\\
& \\
\dfrac{1}{2\pi}\ln\left\vert \mathbf{r}\right\vert  & \text{if }n=2\\
& \\
-\dfrac{1}{\left(  n-2\right)  S_{n-1}}\dfrac{1}{\left\vert \mathbf{r}%
\right\vert ^{n-2}}\ \  & \text{if }n>2
\end{array}
\right.
\]
is Green's function of the lapacian operator in $\mathbb{R}^{n}$.

Representation of the gravitational interaction in $\mathbb{R}^{n}$ via
relation (\ref{ngrav}) is not original but it is presented here in a rational way.

\subsection{Equilibrium}

Let us describe global dynamical properties of a system of $N$ particles of
same\footnote{this assumption is not essential but it simplifies notably the
notations, mainly in the continuous case where $N\rightarrow\infty$} mass $m$
represented by their positions and impulsions merged in the vector$\;\Gamma
_{_{i=1...N}}:=(\mathbf{r}_{i}(t),\mathbf{p}_{i}(t)=m\mathbf{\dot{r}}_{i})$.
Three tensors are of interest : The kinetic $\mathcal{K},$ potential
$\mathcal{U}$ and inertial $\mathcal{I}$ one, which components are
respectively
\[
\forall i,j=1,\cdots,N\ \ \ \ \ \ \ \ \mathcal{K}_{ij}:=\frac{\mathbf{p}%
_{i}\mathbf{p}_{j}}{2m}\;\;\;\;\;\;\mathcal{U}_{ij}:=mGS_{n-1}\sum_{j}%
g_{n}(\mathbf{r}_{i})\;\;\;\;\;\;\mathcal{I}_{ij}:=m\mathbf{r}_{i}%
\mathbf{r}_{j}\text{\ \ .}%
\]
In the continuous case where $N\rightarrow\infty$,
variable is no longer $\Gamma_{i}$ but its probability density
$f=f(\mathbf{\Gamma}_{1},\cdots,\mathbf{\Gamma}_{N},t).$ Dynamical tensors
become usually
\[%
\begin{array}
[c]{ccc}%
\  & \mathcal{K}_{ij}:=%
{\displaystyle\int}
\dfrac{p_{i}p_{j}}{2m}\ f\ d\mathbf{\Gamma}^{N} & \ \ \mathcal{U}_{ij}:=-m%
{\displaystyle\int}
r_{i}\dfrac{\partial\psi}{\partial r_{j}}\ f\ d\mathbf{\Gamma}^{N}\\
\forall i,j=1,\cdots,N &  & \\
& \mathcal{I}_{ij}:=m%
{\displaystyle\int}
r_{i}r_{j}\ f\ d\mathbf{\Gamma}^{N} & \text{where \ }d\mathbf{\Gamma}%
^{N}=\prod\limits_{i=1}^{N}d\mathbf{\Gamma}_{i}%
\end{array}
\]

In the general conservative case $f$ obeys Liouville equation, which reduces
under generic symetry assuptions to Collisionless Boltzmann Equation for
gravitating systems (see \cite{5} for example).

It is not too long so, to prove\footnote{from Newton fundamental principle in
the discrete case or from Liouville equation for the continuous case ...}
\ the fundamental virial theorem

\begin{theorem}
If $\;\mathcal{U}$ is homogeneous of degree $\alpha ,$ i.e. $\forall \lambda
\in \Bbb{R},~\mathcal{U}(\lambda \mathbf{r}_{1},\cdots ,\lambda \mathbf{r}_{N})=$ $\lambda ^{\alpha}%
\mathcal{U}(\mathbf{r}_{1},\cdots ,\mathbf{r}_{N})$ then
\[
2\mathrm{Tr}(\mathcal{K})-\alpha \mathrm{Tr}(\mathcal{U})=\frac{1}{2}\frac{%
d^{2}\mathrm{Tr}\left( \mathcal{I}\right) }{dt^{2}}
\]
\end{theorem}

It is quite natural to define an equilibrium state by
\[
\frac{d^{2}\mathrm{Tr}\left(  \mathcal{I}\right)  }{dt^{2}}=0
\]
in some mean sense. Hence, for a self-gravitating system in a $n-$dimensional
space :

\begin{itemize}
\item If $n=2$, $\mathcal{U}$ is not homogeneous : Virial theorem does not apply
in this form;

\item If $n\neq2$, $g_{n}$ then $\mathcal{U}$ is an homogeneous function of
degree $(n-2)$ and equilibrium is characterized by the relation
\[
2\mathrm{Tr}(\mathcal{K})+(n-2)\mathrm{Tr}(\mathcal{U})=0
\]

\end{itemize}

For extended self-gravitating systems like globular clusters or galaxies a 3
dimensional space in virial theorem seems compatible with observations.

\section{Thermodynamics}

If all particles have the same probability law, are independant and do not
interact by pair but only globally through their whole mean gravitating field, one
can reduce the dimensionality of the phase space : The system is statistically
equivalent to a test particle of mass $m$ at position $\mathbf{r}\in
\mathbb{R}^{n}$, with impulsion $\mathbf{p}\in\mathbb{R}^{n}$, described at
any time $t$ by a distribution function $f(\mathbf{r},\mathbf{p},t)$ and
evolving in a mean field $\psi\left(  \mathbf{r},t\right)  .$ These two
functions are solutions of the Vlasov-Poisson system
\[
\left\{
\begin{array}
[c]{l}%
\dfrac{\partial f}{\partial t}-m\dfrac{\partial f}{\partial\mathbf{p}}%
.\dfrac{\partial\psi}{\partial\mathbf{r}}+\dfrac{\mathbf{p}}{m}.\dfrac
{\partial f}{\partial\mathbf{r}}=0\\
\\
\psi=G\ S_{n-1}\ g_{n}\ast\left[  m%
{\displaystyle\int}
\;f\;\mathrm{d}\mathbf{p}\right]  \mathbf{\;}%
\end{array}
\right.
\]

\subsection{Definitions}

Several quantities are in general, used in thermodynamics :

\begin{itemize}
\item Phase space variable $\mathbf{\Gamma}$%
\[
\mathbf{\Gamma}=\left(  \mathbf{r},\mathbf{p}\right)  \in\mathbb{R}^{n}%
\times\mathbb{R}^{n}%
\]

\item Space ($\nu)$ or mass ($\rho)$ density 
\[
\nu(\mathbf{r},t)=\int f(\mathbf{\Gamma},t)\,\mathrm{d}\mathbf{p}\;=:\rho/m
\]

\item \textquotedblright Number of particle\textquotedblright\ $N$ or system's
mass  $M$%
\[
N[f]:=\int f(\mathbf{\Gamma},t)\,\mathrm{d}\mathbf{\Gamma}=\int\nu
(\mathbf{r},t)\,\mathrm{d}\mathbf{r:=}M/m
\]

\item Energy 
\[
E[f]:=K[f]+U[f]\text{ \ \ \ with \ }\left\vert
\begin{array}
[c]{l}%
K[f]:=\dfrac{1}{2m}%
{\displaystyle\int}
\mathbf{p}^{2}f(\mathbf{\Gamma},t)\,\mathrm{d}\mathbf{\Gamma}\\
\\
U[f]:=\dfrac{m^{2}}{2}%
{\displaystyle\int}
\nu(\mathbf{r},t)\psi(\mathbf{r},t)\,\,\mathrm{d}\mathbf{r}%
\end{array}
\right.
\]

\item Angular Momentum\footnote{the quantity $p_{\phi}$ indicates the
tangential components of the impulsion $\mathbf{p}$, the quantity $r$ stands
for the euclidian norm of the position $\mathbf{r}$ of the test particle.}

\[
J[f]=\int rp_{\phi}f(\mathbf{\Gamma},t)\,\,\mathrm{d}\mathbf{\Gamma}%
\]

\item Boltzmann Entropy 
\[
S[f]:=-\int f(\mathbf{\Gamma},t)\ln[f(\mathbf{\Gamma},t)]\,\,\mathrm{d}%
\mathbf{\Gamma}%
\]

\end{itemize}

For precise considerations let us define some set :

\begin{enumerate}
\item Unbounded Systems
\[
\mathcal{G}_{n}(N,E)\;=\;\left\{  f\;\;\text{\ s. t.}\;\;\forall
\mathbf{\Gamma}\;\;\left\vert
\begin{array}
[c]{lcl}%
K\;<\;\infty & , & U\;<\;\infty\\
E\;<\;\infty & , & N\;<\;\infty
\end{array}
\right.  \right\}
\]%
\[
\mathcal{G}_{n}(N,E,J)\;=\;\mathcal{G}(N,E)\;\cap\;\left\{  f\;\;\text{\ s.
t.}\;\;\forall\mathbf{\Gamma}\;\;\;J\;<\;\infty\right\}
\]

\item Bounded Systems $D\subset R^{n}$

We designe by \textrm{Supp}$(f)$ the support of the distribution function $f$
, i.e. the complementary of the largest open set of $\Gamma\in\left(
\mathbb{R}^{n}\times\mathbb{R}^{n}\right)  $ such that $f\left(
\Gamma\right)  =0$. We then call
\[
\mathcal{G}_{n}(D,N,E)\;=\;\left\{  f\in\mathcal{G}_{n}%
(N,E)\;\;;\;\;\mathrm{Supp}(f)\subset D\right\}
\]%
\[
\mathcal{G}_{n}(D,N,E,J)\;=\;\left\{  f\in\mathcal{G}_{n}%
(N,E,J)\;\;;\;\;\mathrm{Supp}(f)\subset D\right\}
\]

\end{enumerate}

\subsection{Thermodynamical Equilibrium Problem}

The classical thermodynamical equilibrium problem concerns the existence of an
entropy extremalizer. Considering a set $\mathcal{G}_{n}$ of acceptable
distribution functions, it can be posed as :
\[
\exists?\;\;f^{+}\in\mathcal{G}_{n}\;\;\;\;\text{s.t.}\;\;\;\;\forall
f\in\mathcal{G}_{n}\;\;S[f]\;\leq\;S[f^{+}]
\]
For sets considered in the previous section, this problem corresponds to a
classical Euler-Lagrange one, which solutions are the well-known isothermal
spheres :

\begin{itemize}
\item For $f\;\in\;\mathcal{G}_{n}(E,N)$
\[
f^{+}=exp\left\{  -\alpha-\beta(\frac{\mathbf{p}^{2}}{2m}+m\psi^{+})\right\}
\]
Lagrange multiplier $\alpha$ and $\beta$ correspond respectively to the
conservation of $N$ and $E$ .
\end{itemize}

\begin{itemize}
\item For $f\;\in\;\mathcal{G}_{n}(E,N,J)$
\[
f^{+}=\exp\left\{  -\alpha-\beta\left(  \frac{{p}_{r}^{2}}{2m}+\frac{(p_{\phi
}-mr\omega)^{2}}{2m}+m\psi^{+}-\frac{m^{2}\omega^{2}r^{2}}{2}\right)
\right\}
\]
The additional constraint of $J$ conservation corresponds to the introduction
of the $\omega$ multiplier.
\end{itemize}

\subsection{$2D$ Thermodynamics : $\mathcal{G}_{2}$}

We reproduce here the classical results obtained in \cite{1} and \cite{2}. The
main one is twofold : We first produce a bound for
entropy, we then study the existence of a distribution function which allow 
reaching that bound. Concerning the upper bound for the entropy, we prove
that\footnote{in the units of original papers}

\begin{theorem}
$\forall f\in \mathcal{G}_{2}(E,N)$ \label{th2}
\[
S[f]\leq S^{+}(N,E):=\sup_{\mathcal{G}_{2}(N,E)}S[f]\leq \frac{2E}{N}+N\ln
(e\pi ^{2})
\]
\end{theorem}

In $2D,$ entropy of unbounded self-gravitating systems for which $E=cst$, and
$N=cst$ is bounded from bellow.

\begin{theorem}
In the notations of Theorem \ref{th2}
\[
S^{+}(N,E,J)=S^{+}(N,E)
\]
\end{theorem}

Adding the angular momentum's constraint does not change the least upper bound
on the entropy. The existence of a distribution function $f^{+}$
corresponding to this entropy maximizer, i.e. $S\left[  f^{+}\right]  $
$=S^{+}$ is closely related to the set under consideration :

\begin{itemize}
\item In $\mathcal{G}_{2}(E,N)$ : $f^{+}$ exists and is unique,
\[
f^{+}=\frac{e^{2(E-N^{2})/N^{2}}}{\pi^{2}}\frac{e^{-p^{2}/N}}{(e^{2(E-N^{2}%
)/N^{2}}+r^{2})^{2}}%
\]
it generates the potential
\[
\psi^{+}=N\ln\left(  e^{2(E-N^{2})/N^{2}}+r^{2}\right)  \,.
\]
in the units of papers \cite{1} and \cite{2}.

\item There is no $f^{+}\in\mathcal{G}_{2}(E,N,J)$ for which $S[f^{+}]=S^{+} $.

\item In $\mathcal{G}_{2}(D,E,N)$ and $\mathcal{G}_{2}(D,E,N,J)$ : $f^{+}$
exists and is unique
\end{itemize}

\subsection{$3D$ Thermodynamics : $\mathcal{G}_{3}$}

Introducing a new dimension changes drastically the situation concerning the
classical equilibrium problem of gravitational thermodynamics. As a matter of
fact, it is well known for a long time (see for example \cite{5} p. 268) that

\begin{itemize}
\item Entropy has no global maximum on $\mathcal{G}_{3}(E,N)$, $\mathcal{G}%
_{3}(E,N,J)$, $\mathcal{G}_{3}(D,E,N)$ and $\mathcal{G}_{3}(D,E,N,J)$.

\item Entropy has no local maximum on $\mathcal{G}_{3}(E,N)$ and
$\mathcal{G}_{3}(E,N,J)$.
\end{itemize}

The existence of local maximum for entropy in $\mathcal{G}_{3}(D,E,N)$
corresponds to an extensive literature initiated by the works of V.A.\ Antonov (see \cite{5} for a review) in the early 60's. It represents a beautifull problem of thermodynamics.

\subsubsection{Entropy extremalizer in $\mathcal{G}_{3}(D,E,N)$}

\begin{center}
\begin{figure}[h]
\centering
\includegraphics[height=7.0468cm,width=7.0951cm] {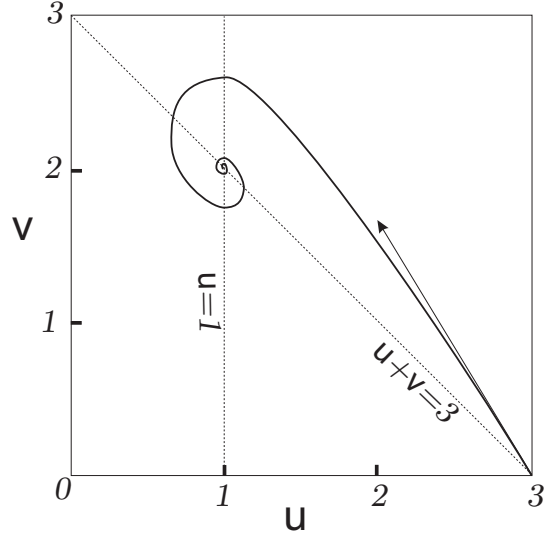}
\caption{Isothermal sphere in the Milne plane}
\label{ant1}
\end{figure}
\end{center}

We reproduce below the main results obtained by T. Padmanabhan (see \cite{3})
which clarify all previous works. If we denote by $R$ the radius of the
largest bowl contained in the spatial part of $D$. It is proven that any
entropy extremalizer in $\mathcal{G}_{3}(D,E,N)$ must be of the form
\begin{equation}
f^{+}=\left(  \frac{2\pi}{\beta}\right)  ^{-3/2}\nu_{o}\;e^{-\beta
E}\;\;\;\text{with}\;\;\;\;\;\;m\nu_{o}=\rho(0)\;e^{\beta\psi(0)}\ .
\end{equation}
The associated potential verifies Poisson equation
\begin{equation}
\Delta_{3}\;\psi=4\pi Gm\nu_{o}\;e^{-\beta\psi}%
\end{equation}
with the limit condition $\psi\left(  R\right)  =-GM/R$.\ Introducing
dimensionless variables%
\[%
\begin{array}
[c]{l}%
L_{o}=\sqrt{4\pi G\rho\left(  0\right)  \beta}\;\;\;\;\;\;M_{o}=4\pi
\rho\left(  0\right)  L_{o}^{3}\;\;\;\;\;\;\psi_{o}=\beta^{-1}=\dfrac{GM_{o}%
}{L_{o}}\\
x=r/L_{o}\;\;\;\;\;\;n=\rho/\rho\left(  0\right)  \;\;\;\;\;\;\mu=M\left(
r\right)  /M_{o}\;\;\;\;\;\;y=\beta\left(  \psi-\psi\left(  0\right)  \right)
\end{array}
\]
Poisson equation becomes
\[
\frac{1}{x^{2}}\frac{d}{dx}\left(  x^{2}\frac{dy}{dx}\right)  =e^{-y}%
\;\;\;\text{with }\;\;\;y\left(  0\right)  =y^{\prime}\left(  0\right)  =0
\]
Milne's functions, $v=\mu/x$ and $u=nx^{3}/\mu,$ transform Poisson equation into
\[
\frac{u}{v}\frac{dv}{du}{=}\frac{1-u}{u+v-3}\;\;\text{with }\;\;\left\vert
\begin{array}
[c]{l}%
v=0~\ \text{when }u=3\\
\ \ \ \text{and}\\
\left.  \dfrac{dv}{du}\right\vert _{\left(  u,v\right)  =\left(  3,0\right)
}=-5/3
\end{array}
\right.
\]
$\;$Hence, Isothermal extremal spheres lie on a curve in the $u-v$ plane.
This curve is plotted on Figure \ref{ant1}.

\subsubsection{Antonov Instability}

\begin{center}
\begin{figure}[h]
\centering
\includegraphics[height=7.0709cm,width=7.0951cm] {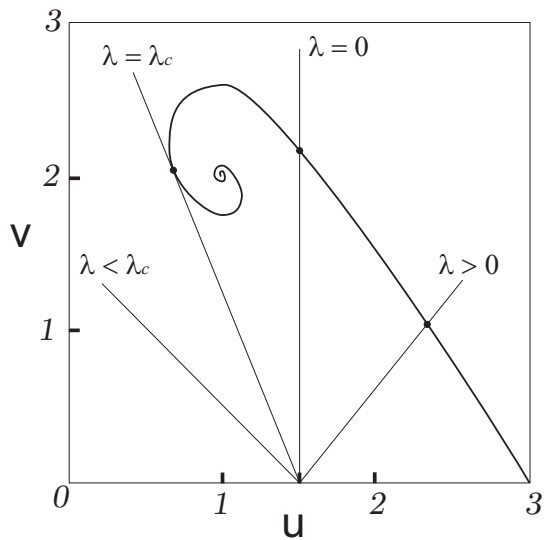}
\caption{Existence of an extremal isothermal sphere}
\label{ant2}
\end{figure}
\end{center}

In a meaningfull remark, Patmanabhan notes that dimensionless quantity
\[
\lambda:=\frac{RE}{GM^{2}}=\frac{1}{v}\left(  u-\frac{3}{2}\right)
\]
lies also on the same $u-v$ plane. He then asks the fundamental question : Can
$E,R$ and $M$ be accomodated by a suitable choice of $\rho\left(  0\right)  $
and $\beta$ ? As one can see on Figure \ref{ant2}, the answer is clearly
no. There exists a critical value $\lambda_{c}\simeq-0.335$, associated with the
possibility to put a given isothermal extremal sphere in a given box !

\begin{itemize}
\item If $\lambda<\lambda_{c}$ isothermal sphere cannot exist, entropy
extremum cannot exist;

\item If $\lambda>\lambda_{c}$ isothermal sphere exist, entropy extremum exist !
\end{itemize}

More exciting is the fact that the nature of extremum depends on
$\kappa=\rho\left(  0\right)  /\rho\left(  R\right)  $ the density concentration ratio
of the isothermal sphere.

\begin{itemize}
\item If $\kappa>709$ : The extremum of the entropy is an unstable saddle point;

\item If $\kappa<709$ : The extremum of the entropy is local maximum.
\end{itemize}

This last point is generally associated to the so-called gravothermal
catastrophe, we prefer to call it Antonov Instability. As we will see in the next
section, such an instability is certainly at the origin of some important
characteristics of extended self gravitating systems.

\section{Antonov instability in astrophysics}

\subsection{Globular clusters in galaxies}

Since early 80's observations have shown that galactic globular clusters split
in two categories which differ by some properties of their radial density
profiles. On the one hand a large familly of about 120 clusters with a large
constant density core which extends to almost the half mass radius ($R_{50}$)
of the whole system. This large core is surrounded by a power law density
decreasing halo. On the other hand, a small familly of about 20 core collapsed
clusters with a very high central density which decreases monotonically outward
with mainly two power law indexes. These two types of globular clusters are
very well represented by two of their components, namely NGC\ 6388 for core
halo cluster and Trz2 for core collapsed cluster, see Figure 1 of \cite{6}.

\begin{center}
\begin{figure}[h]
\centering
\includegraphics[width=7.0951cm] {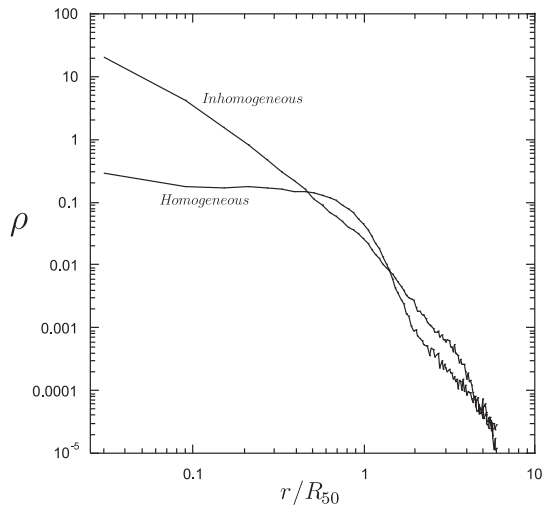}
\caption{Radial density profile obtained from the gravitational collapse of an
homogeneous set of mass and an inhomogeneous one}
\label{retp}
\end{figure}
\end{center}

Such a behaviour of the radial density could be explained in a very simple way
by Antonov Instability. As a matter of fact, if globular cluster formation results from
the collapse of a small, hence homogeneous, region of some galaxy, the natural
result is roughly an isothermal sphere (see \cite{4}) with generally a
contrast density $\kappa$ less than the critical value. The evolution of the
cluster in the galaxy produces a slow evaporation of the cluster (passing
through the galalactic plane in spiral galaxies for example). Due to the
negative specific heat of such gravitating systems, this evaporation makes the
contrast density growing. When $\kappa$ reaches the critical value, Antonov
instability triggers and transforms the core halo density profile into a
collapsed core one. On Figure \ref{retp} we represent radial density profile after
collapse of an initially homogenenous system and of an initially inhomogeneous
one (see \cite{4}).

\subsection{ A paradigm for Super Massive Black Hole(SMBH) formation}

From the accumulation of observational data, it becomes necessary to put a
Super Massive (from $10^{6}$ to perhaps $10^{9}$ $M_{\odot}$) Black Hole in the dynamical center
of galaxies.  Excepted the fact that such gravitational monsters are as older as
their host, little is known about their formation process. In the context of
hierachical galaxy formation scenario, Antonov instability could produce a
good paradigm. As a matter of fact, if galaxies are the net result of
successive collapse and merger of gravitational structures, it can be
modelized generically by a general collapse of inhomogeneous media. As showed
by \cite{4}, in such case, reaching the center small structures first collapse
to form quasi-isothermal sphere surrounded by the rest of not yet collapsed
large structures. Evaporating the smallest, largest's collapse could trigger
Antonov instability. Progenitor of  SMBH could then be formed. This process is
up to now to be confirmed but seems correspond to all observed properties.

\section{Conclusion}

Classical gravity is an amazing topic. Although, $2D$\ gravitating systems are
well described by thermodynamics, their equilibrium is not well defined. By
opposition, provided that $n\geq3$, we possess a powerfull tool to describe
gravitational equilibrium for systems in $\mathbb{R}^{n}$, but the
corresponding thermodynamics is not too efficient. However, in the restricted
case of bounded systems in $\mathbb{R}^{3}$, the message from gravitational
thermodynamics and particulary Antonov instability, could be fundamental
to explain some features of self gravitating systems.

\end{document}